\documentclass[journal=jpcafh,manuscript=letter]{achemso}
\usepackage[version=3]{mhchem}

\author{Seyed Mohammad Ghazi}
\email{mohammad@physics.unipune.ernet.in}
\altaffiliation{Ministry of Science, Research 
and Technology, Tehran , Iran}
\author{Shahab Zorriasatein}
\email{shahab@unipune.ernet.in}
\author{D. G. Kanhere}
\email{kanhere@unipune.ernet.in}
\affiliation[Pune University]
{%
Department of Physics and
Center for Modeling and Simulation,
University of Pune,
Pune 411 007,
India.}

\title{%
Building clusters atom by atom: from local order to global order. 
}

\begin{document}

\begin{abstract}
We have carried out extensive density functional calculations for series
of sodium clusters $Na$$_{N}$ ranging from $N$=10 to 147 and have 
obtained $\approx$ 13000 distinct isomers. We unravel a
number of striking features of the growth characteristics. The growth
shows order-disorder-order pattern of cyclic nature. Between two ordered
clusters the growth proceeds via disordered clusters having multi-centered
icosahedral local order. 
The Global order emerges suddenly with the addition of one or
two atoms only. The clusters around $N$=92, the electronically 
closed shell system, behave completely differently and do not 
show the favored icosahedral
local order. It is the absence of icosahedral local order which is responsible
for rather low melting temperatures observed in the experiments.
\end{abstract}

Clusters are a focus of considerable 
attention due to their importance as building blocks for nano materials 
as well as due to intrinsic properties arising out of their finite size. 
It is now well established that in the small  
size regime, typically below few 
100 atoms, clusters show individuality. Specifically, they show size 
dependent 
features in their geometries, energy gaps and binding energies, stability,
polarizability,~\cite{francesca, jena1, kumar1, haber-book, 
jellinek, kronik}  
melting temperatures, and 
shapes of the heat capacity curves~\cite{haberland, jacks, breaux, aguado1}.  
In spite of a large number of studies  
a clear evolutionary pattern of the growth over a wide range of 
size has not been developed. There is no clear answer to a rather 
simple question of interest 
namely how does a cluster grow atom by atom? 
Most of the systematic theoretical studies at {\it ab initio} level are limited to 
sizes below 40 atoms or so and describe
geometries of individual clusters and energetics. 
An interesting work explaining shape transition in 
$Si_{\rm N}$ ($N$=20-27) has been reported by Jackson {\it et al.}~\cite{jackson1}.

Clusters of sodium atoms are perhaps the most intensively 
studied both experimentally
as well as theoretically~\cite{haberland, kostko, calvo, manien, kumar2, ursella, ghazi}.
Although features like stability can be understood on the basis of 
simple jellium models, {\it ab initio} methods are required for determining
the geometries and other properties. 
Even for these simple metal atom systems there is no  
understanding of the evolutionary behavior. These clusters 
also show very peculiar irregular behavior in some of the properties
like melting temperature~\cite{haberland}. 
Most of the early work on these clusters is restricted to $N$ $<$ 20. 
For example R{\"o}thlisberger et al.~\cite{ursella} have 
discussed  the nature of individual geometries 
and possible occurrence of some motifs as atoms are added one by one.

In the present work the focus is 
on understanding the systematics of structural evolution of $Na$$_{N}$ over a 
wide size range: $N$=10-147. We examine the basic question;  
how do sodium clusters grow when a single atom is added 
starting from $N$=10? 
We have carried out extensive density functional calculations
and have obtained at least 200 distinct isomers 
for all the clusters for $N \leq 80$ and at least 100 isomers 
for all the clusters 
having even number of atoms with 80 $<$ $N$ $ \leq 147$.   
All the calculations have been performed within unified framework using
plane wave basis, ultrasoft
pseudopotential, local density exchange correlation potential of
Ceperley-Alder, 
and the same box length(36 \AA)~\cite{vasp}.   
Our results are based on the analysis of about 13000
equilibrium structures for 105 clusters. 
These equilibrium structures were obtained by minimizing a 
few hundred (100-200)
initial configurations per cluster. These configurations were chosen 
from ab initio constant temperature
molecular dynamical runs carried out at 2 different temperatures 
for the period of 50 to 90 ps. 
All the lowest energy structures are subjected to vibrational analysis.
For many of the large clusters nearly degenerate 
levels near Fermi level were handled by 
redistributing the occupancies by assigning appropriate low temperature to the 
electrons.

We have analyzed the ground-states structures 
mainly focusing on the nature of their geometries.
We examine their shapes by using shape deformation parameter 
($\varepsilon_{\mathrm{def}}$) defined below, 
the geometric shell 
structure as viewed from the 
Center Of Mass(COM) and the distribution of all the relevant bond length
in the system. 
 $\varepsilon_{\mathrm{def}}$ is defined as:
$\varepsilon_{\mathrm{def}} = \frac{2Q_{x}}{Q_y+Q_z}$, where 
$Q_x \geq Q_y \geq Q_z$ are the eigenvalues 
of the quadrupole tensor $Q_{ij} =  {\sum_{I}  R_{Ii}\,R_{Ij} }$.
Here $R_{Ii}$ is the $i$th coordinate of ion $I$ relative
to the COM.
We also examine the coordination numbers
for all the atoms to locate possible existence of any local order.
Then we examine the motifs formed by $\approx$12 nearest neighbors for
all the atoms having 11-13 nearest neighbors.
We also calculate the surface energy $S$ 
through the following relationship:
$S= E_{\rm tot}(N)-\epsilon_{\rm \infty}N$
where
$E_{\rm tot}(N)$ is the total energy of the cluster and 
$\epsilon_{\rm \infty}$ is the energy per atom of the bulk sodium.
The surface area of a given cluster is calculated by 
approximating the shape as ellipsoidal~\cite{elipse}.

It is fruitful to begin by noting some characteristic features
in the geometries seen over the entire series. The series has two 
distinctive clusters
$Na$$_{55}$ and $Na$$_{147}$ which are complete McKay icosahedra 
displaying geometric shell
closing and a full five fold rotational symmetry. 
$Na$$_{13}$, a possible icosahedral
structure is seriously distorted due to 
Jahn-Teller distortion. We call the two 
above clusters as completely ordered. Apart from these two spherically 
symmetric clusters, the clusters with $N$=20,40,58,92, and 138 are electronically
closed shell systems and are also spherical. We also note that among
these, $Na$$_{92}$ is very special in the sense that its location is far from
complete icosahedral clusters namely $Na$$_{55}$ and $Na$$_{147}$. As  
we shall see
the atomic arrangement in the clusters 
around $N$=92 will turn out to be dramatically 
different than the growth pattern observed after $N$ $>$ 55. 

Although the general principle determining the shapes of these clusters
is the total energy minimization its manifestation on the growth pattern 
is due to a delicate balance between two competing energies: the surface energy 
which will tend to make the cluster spherical and  
the nature of the binding  
which would like to place the atoms so as to 
have optimum coordination number 
to maximize the binding energy locally.
The interesting issues we address are: (1) how do 
cluster grow from one ordered cluster
to another ordered cluster? (2) How do the competing 
interactions noted above influence
the growth pattern locally as well as globally? 
(3) Is there a characteristic difference
between the atomic arrangements of geometrically closed 
shell clusters and electronically closed shell
clusters?    

Our analysis uncovers a number of striking 
features of the growth characteristics. 
Firstly, the growth shows order-disorder-order pattern of cyclic nature. 
Secondly, we observe  
firm local order, in this case icosahedral local order as the growth 
proceeds between the two ordered structures.  
Thirdly, we also observe a peculiar atomic arrangement in the clusters 
around $N$=92. In fact it turns out that these clusters do not
exhibit the favored icosahedral local order.  

\begin{figure}
\begin{center}
\centerline{\includegraphics[width=0.72\textwidth]{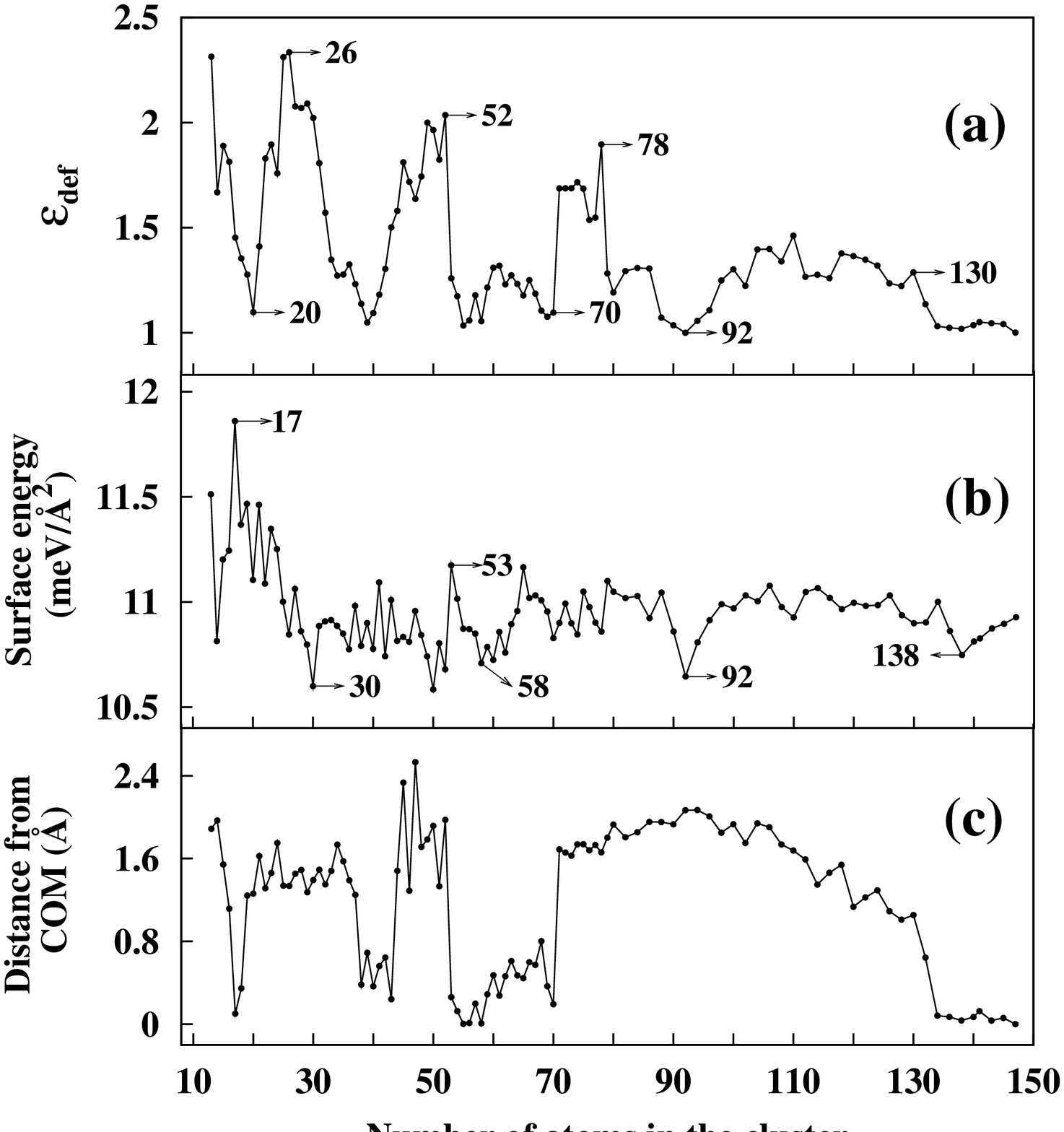}}
  \caption{}
  \label{eps_def.Na12-147.eps} 
  \end{center}
\end{figure}

\begin{figure}
\begin{center}
\centerline{\includegraphics[width=.65\textwidth]{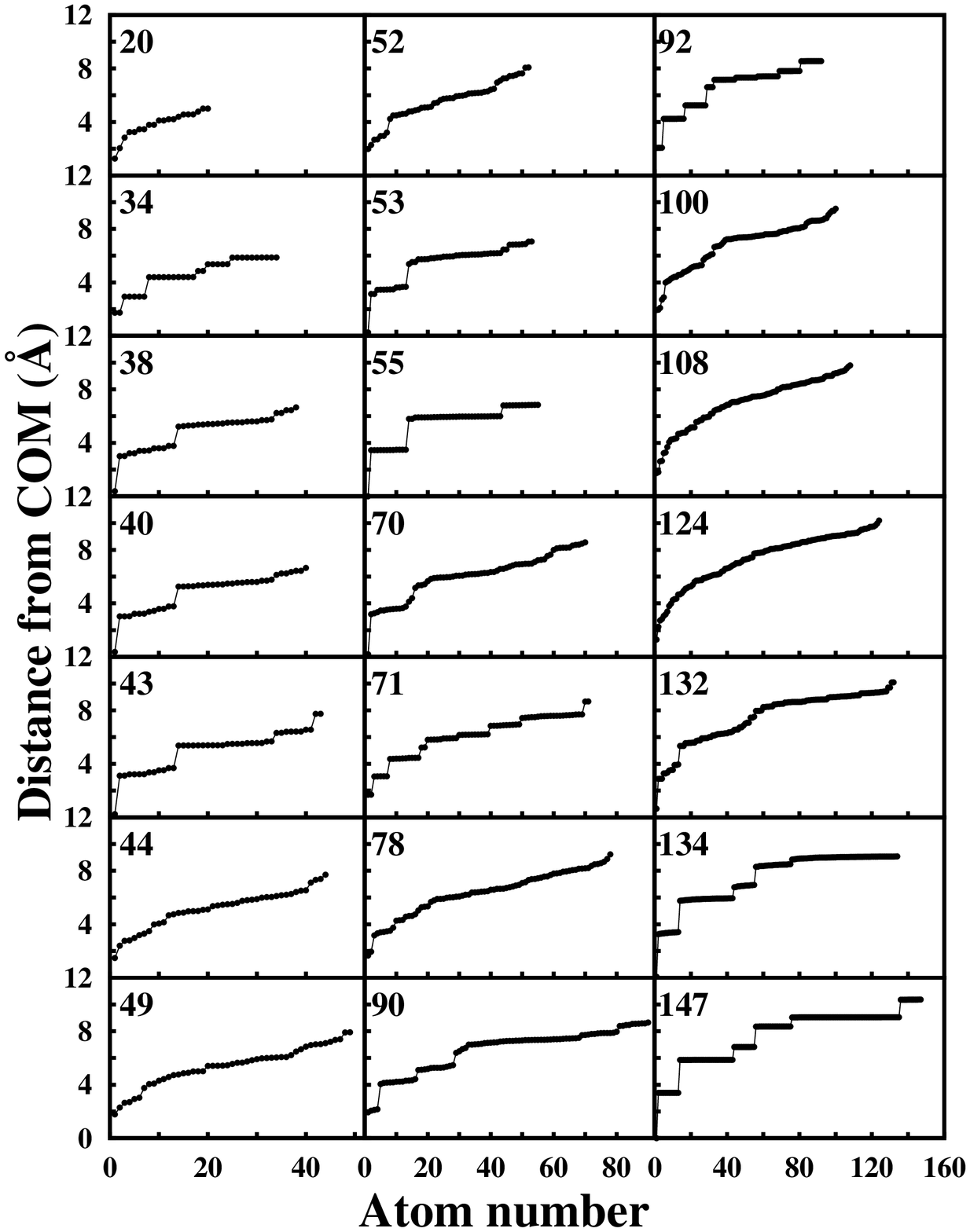}}
  \caption{}
  \label{dist_from_com5.eps} 
\end{center}
\end{figure}

We begin the discussion by presenting in \ref{eps_def.Na12-147.eps},
$\varepsilon_{\mathrm{def}}$ 
, the surface 
energy per unit 
area and the distance of the nearest atom from the COM 
for the entire series.
There are a number of interesting features that are seen in the figure.
The shape changes cyclically from spherical
 ($\varepsilon_{\mathrm{def}}$ $\approx$ 1) 
to the non spherical ($\varepsilon_{\mathrm{def}}$ $\approx$ 1.8) 
and back to spherical structure.
Thus, clusters with
 $N$=20,40,55,70,92 and within the range of 
134-147 are spherical. 
The change of the shape from non spherical to spherical is rather abrupt
as indicated by sharp drops in $\varepsilon_{\mathrm{def}}$, 
normally by addition of two to three atoms, $N$=92
being an exception.  
The minima and the maxima in the shape parameter correlates
extremely well with the behavior of surface energy, it being minimum for
spherical systems. Interestingly for $ N \leq 70$ the 
spherical clusters also have an atom very near or on the COM.
However from $N$=71-108 i.e over a wide range of sizes, the nearest atom 
from the COM is away at about 1.5 \AA. This happens to be true 
even for $Na$$_{92}$, a spherical cluster.
We have carefully examined the detailed arrangement of atoms in the GS of
all the clusters by calculating the distance of all the atoms from 
the COM.
In \ref{dist_from_com5.eps} we show the distance 
from the COM, ordered in 
increasing fashion, for all the atoms. Such a plot shows characteristic
step like structure if the cluster contains a well defined shell structure.
For example $Na$$_{147}$ shows characteristic steps containing
13,30,12,20,60, and 12 number of atoms corresponding to complete icosahedron.
The representative samples shown in \ref{dist_from_com5.eps} 
is large enough to show the 
formation and destruction of these shells as the clusters grow. 
It can be seen that 
for the clusters in the range $N$=19-37 there is hardly any 
shell formation(except at $Na$$_{34}$).
Although shell formation is seen around $N$=40, it gets destroyed at $N$=44
and seen again at $N$=53 onwards. Again clusters in the range of $67 \leq$ $N$ $<$ 134  
have no obvious shell structure except at $N$=92. %80 $<$ $N$ $ \leq 147$
All the clusters having high value of $\varepsilon_{\mathrm{def}}$
do not show any shell structure, have higher surface energy per unit
area and also do not have an atom at or near the COM. 
A detailed examination of the atomic arrangements reveals that there is no
particular rotational symmetries present in these clusters. 
Thus, these clusters are disordered. 

\begin{figure}
\begin{center}
\centerline{\includegraphics[width=1.2\textwidth]{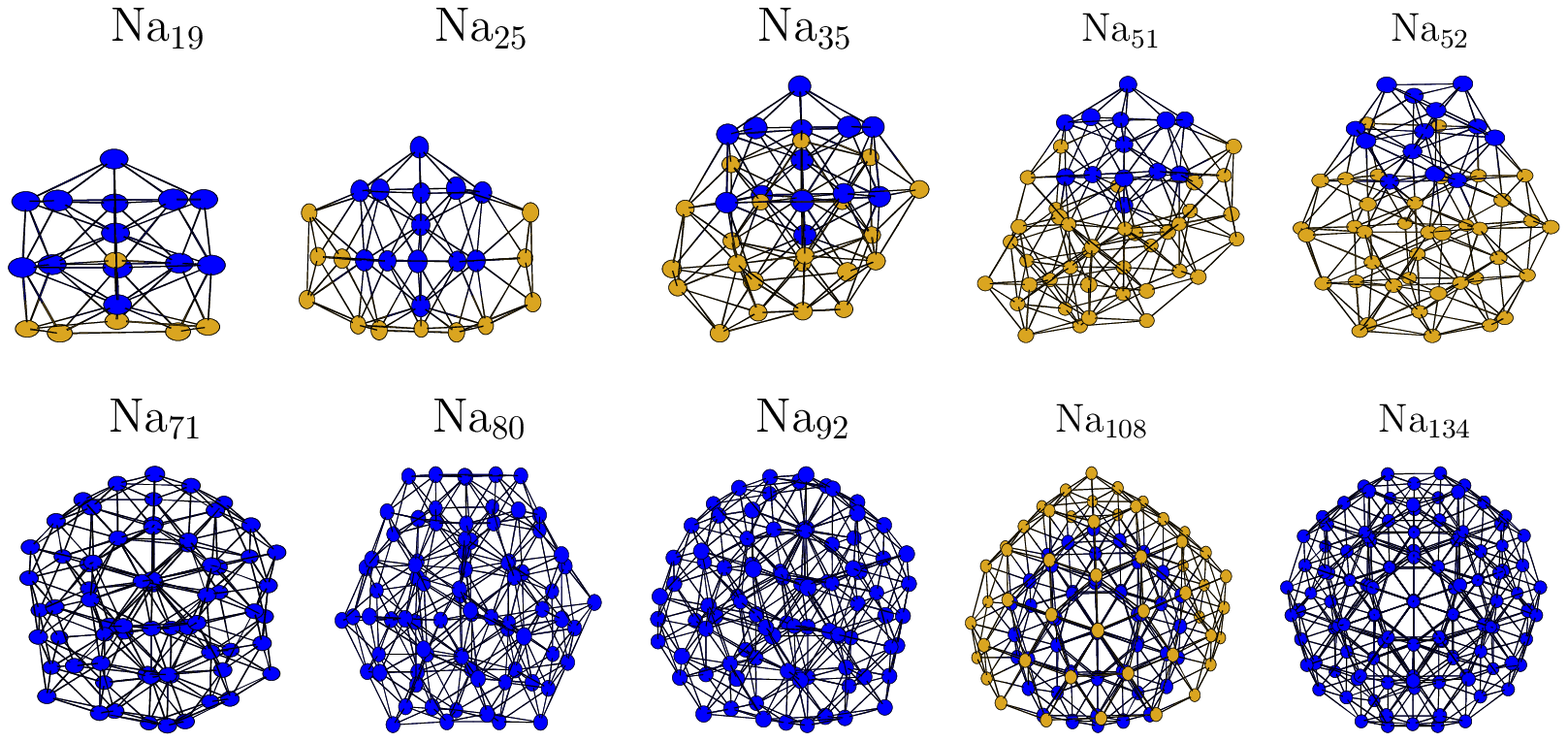}}
  \caption{}
  \label{GS_geo.ps}
\end{center}
\end{figure}

It turns out that the clusters grow from one ordered 
geometry to another one via a disordered growth. 
These disordered clusters show large deviations from the 
sphericity. The exception is $N$=92 the electronically 
driven system which we shall
discuss separately.
We have carefully examined the coordination numbers of all the atoms in each of the 
cluster. Then we focused on the atomic 
arrangements around all the atoms in a given clusters
having 12 nearest neighbors.
Strikingly, the thirteen atom motif thus formed turns out to be that of {\it icosahedron}. 
In fact there is a pattern in the formation of such centers of local order 
in these disordered clusters.  
As the clusters grow atom by atom, 
the number of such centers grow forming interpenetrating icosahedra. 
The first appearance of icosahedron occurs 
at $N$=19.  \ref{GS_geo.ps} shows some representative 
geometries between $N$=19 and
$N$=134 where we have highlighted at least one such icosahedron. 
The number of such centers grow to about 8 at $N$=40 and to about 10 at $N$=52.    
The emergence of the global order out of such multi-centers of local 
order can best be illustrated by examining the growth from $N$=45 to $N$=55. 
Let us recall that the ordered cluster $Na$$_{55}$ 
consist of one central 13 atom 
icosahedron and has 12 peripheral decahedra.
As the clusters grow from $N$=40  
they 
become progressively disordered, their shell structure
is destroyed, and $\varepsilon_{\mathrm{def}}$  
grows to $\approx$2
till $N$=52. With the addition of one atom this cluster completely reorganizes 
into nearly spherical, well-ordered icosahedron with 2 atoms missing from 
the last shell. This transformation is sharp and is driven by 
surface energy. Thus the global order emerges out of locally ordered 
but globally disordered clusters rather suddenly.   
A similar pattern is observed as the
clusters grow after $N$=55.
We observe one icosahedral center having 55 atoms between $N$=56-70.
At $N$=71
two interpenetrating icosahedral motifs appear. This is accompanied by a sharp 
increase in $\varepsilon_{\mathrm{def}}$, accompanied by displacement of the atom
on the COM by about 1.5\AA~ and the cluster becomes prolate. 

\begin{figure}
\begin{center}
\centerline{\includegraphics[width=.9\textwidth]{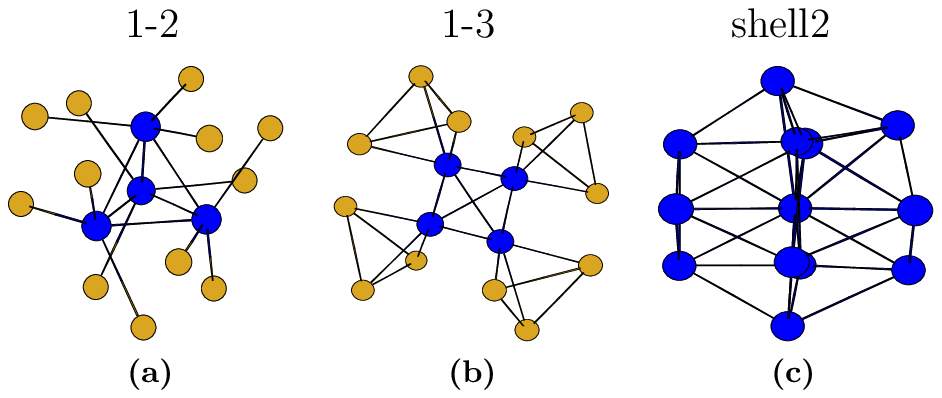}}
  \caption{}
   \label{motif_92.eps}
\end{center}
\end{figure}

A third such center appears at $N$=80
making the cluster($Na$$_{80}$) more spherical.     
This pattern of forming local icosahedral order with 55 atoms
is interrupted around $N$=88 and is established again at $N$=108.
It may be noted that $Na$$_{92}$ is an electronically closed shell system,
having spherically symmetric charge density, forcing the ionic 
geometry to be spherical. The shell 
structure evident in \ref{dist_from_com5.eps} is not
that of icosahedron.
There is no rotational symmetry in the system.
However there is a peculiar local order that can be discerned.
In \ref{motif_92.eps}(a), \ref{motif_92.eps}(b) 
we show all the atoms in the first 2 shells and 
in the first and third shell 
respectively. The rotational symmetry in the motifs shown is 
evident. Thus, this globally disordered cluster shows strong
local order although not icosahedral. 
All the atoms in the first three shells have 12 
nearest neighbors. Interestingly within each of the shells the 12 atom 
motif formed by these nearest neighbors is unique and is 
identical(up to 2 decimal places) for all the atoms in that shell.
There are no icosahedral motif seen in the cluster but we do see around 
third shell atoms decahedral motif.
A typical one seen around the second shell atoms is 
shown in \ref{motif_92.eps}(c).
The motif 
around the third shell atom is slightly distorted decahedron(not shown). 
The expected
growth pattern for $N$$>$88 should have given rise to interpenetrating 
multi-centers of 55 atoms icosahedra. Instead, 
this electronically driven spherically symmetric cluster 
develops a peculiar local order which is not icosahedral.
The 55 atom icosahedron reappears at $N$=108 but is not centered on COM.
Evidently (\ref{dist_from_com5.eps}) the icosahedral shell 
structure as seen from the COM is completely
recovered at $Na$$_{134}$.
A careful examination of \ref{eps_def.Na12-147.eps}(c) 
and  \ref{dist_from_com5.eps}
for clusters in the range $N$=130 to 134 clearly bring out 
the sharp effect of adding one or two atoms.
All the atoms after $N$$>$134 just fill in the last shell
without destroying the shape. 

Recently Kostco {\it et al.}~\cite{kostko}  
reported ground state structures of 
sodium clusters using photoelectron spectroscopy. They conclude that 
the structures at intermediate sizes between closed shell icosahedra
($N$=55-147) are formed by growth of layers on icosahedral motifs.
Our geometries for  $Na$$_{71}$ and $138 \leq $N$ \leq 147$ are 
consistent with this. However
the clusters around $Na$$_{92}$ 
(especially 86 $<$ $N$  $<$  108) do not show any
icosahedral motif.

Finally we offer a clear physical explanation for the most
important experimentally observed feature 
namely rather low values of 
melting temperature(210 K) around $N$=92~\cite{haberland}. 
Quite clearly the clusters in this region are characterized by
absence of local icosahedral order. 
In fact the bond length of the 12 nearest neighbors atoms 
from the central one forming the motif discussed above differ
by as much as 0.5 \AA.
This is in sharp contrast to the other clusters in the series.
It is precisely the absence of local icosahedral order which
is responsible for rather low melting points. 
The sharp rise in the melting temperature seen around $N$=134 coincides 
with the establishment of central
icosahedron consisting of well-formed 5 icosahedral shells.  

In conclusion the growth of sodium clusters $Na_{\rm N}$ ($N$=10-147) shows an 
order-disorder-order pattern. We observe formation of multi-centers
having strong icosahedral local order specially in the disordered clusters.
The establishment of global order is sudden, accompanied by a sharp change in the shape 
parameter. Clusters around $N$=92(electronically closed shell system) show 
no local icosahedral order. We attribute observed low 
melting temperatures to the absence of favored icosahedral order.    

\acknowledgement  
One of us
(S. M. Ghazi) acknowledges financial support from the Indian Council for
Cultural Relations (ICCR).

\newpage

{\huge \bf Captions for Figures} 

\vskip 0.3in

fig.1.    (a) The shape deformation parameter, (b) surface energy per unit area, (c) distance of the nearest atom from the COM as a function of size for the $Na_{\rm N}$ $N$=13-147.

fig.2.    The distance from the COM for each atom ordered in the increasing fashion for the ground-states of selected clusters. The sharp steps indicate formation of geometric shells.

fig.3.   (Color online). The ground-state geometries of some clusters. Motifs formed by dark spheres(blue online) show icosahedral local order.

fig.4.   (Color online).(a) and (b) Atomic arrangement for the first two and the first and third shells in $Na$$_{92}$. (c)The unique motif seen around all the atoms in the second shell of $Na$$_{92}$.

fig.5.    (Color online). Ground state geometry of Na$_{108}$ depicting 
                          formation of icosahedra. (Blue spheres) 
\newpage

\newpage

\newpage

\newpage

\newpage

\begin{figure}
\begin{center}
\centerline{\includegraphics[width=.9\textwidth]{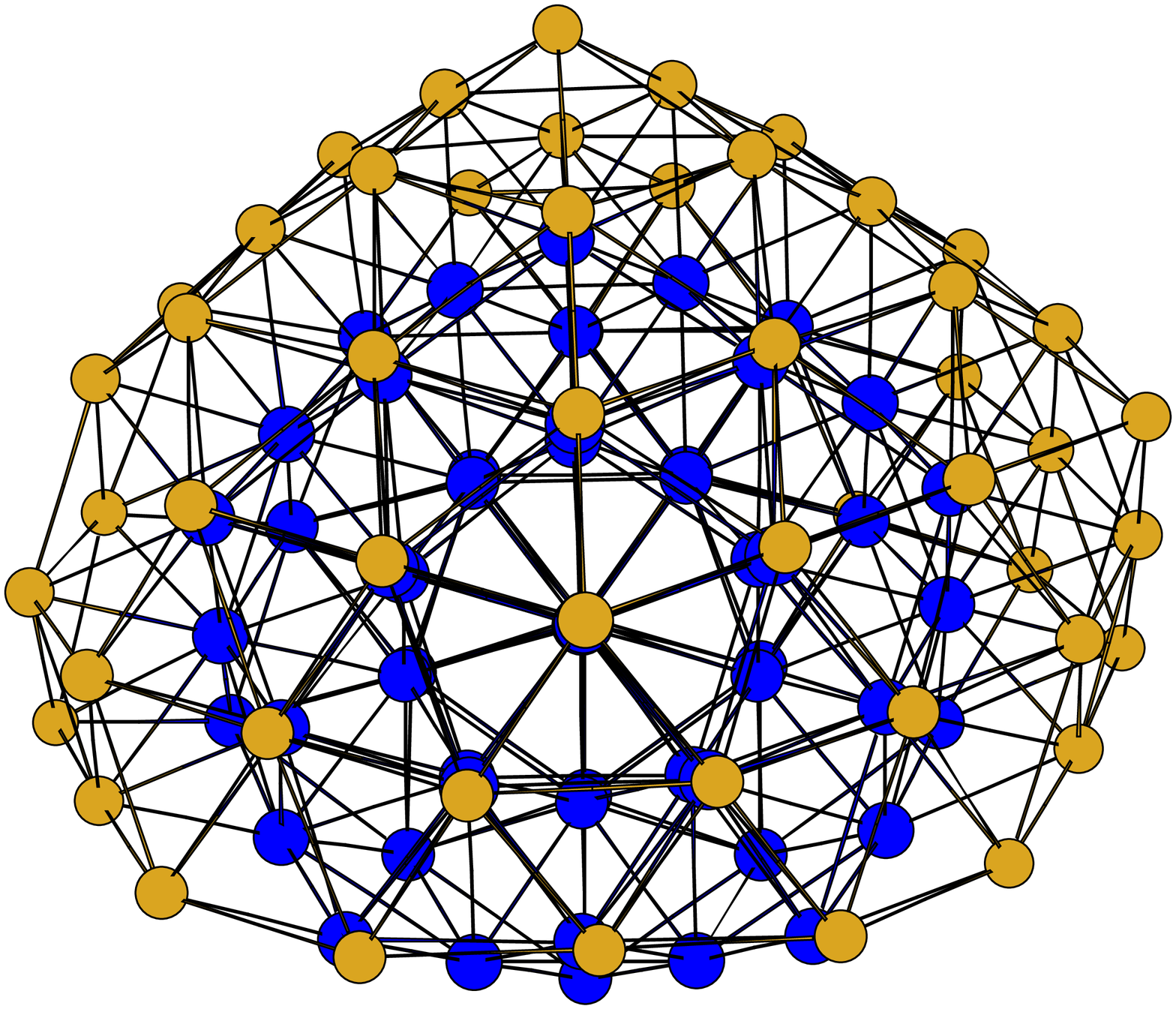}}
  \caption{}
   \label{Gs.108.mot1.eps}
\end{center}
\end{figure}


\begin{thebibliography}{}
\bibitem{elipse}      The surface area $s$ is calculated via 
             $s\approx4\pi(\frac{a^pb^p + a^pc^p + b^pc^p}{3})^{\frac{1}{p}}$
             where $a$, $b$, and $c$ are elliptic radii and p$\approx$1.6075.
\end{thebibliography}
\end{document}